\documentclass[aps,prl,twocolumn,showpacs]{revtex4-1}
\usepackage{graphicx}
\usepackage{bm}
\usepackage{wasysym}
\usepackage{setspace}
\usepackage{mathrsfs}

\begin{document}

\title{Fluctuations in Strongly Correlated Electron Systems}
\author{Mu-Kun Lee}
\author{Tsung-Sheng Huang}
\author{Chyh-Hong Chern}
\email{chchern@ntu.edu.tw} 
\email{chern@alumni.stanford.edu}
\affiliation{Department of Physics, National Taiwan University, Taipei 10617, Taiwan}
\date{\today}
\begin{abstract}
High transition temperature superconductors in cuprates exhibit the charge-density-wave fluctuations and the ferromagnetic time-reversal-symmetry-breaking fluctuation in the polar Kerr rotation experiments.  We demonstrate that they share the same root of origin, and the underlying mechanism also leads to the pseudogap formation.  The pseudogap formation, the charge-density-wave fluctuation, and the time-reversal-symmetry-breaking fluctuation are the consequent phenomena of the correlation.  They are the basic notions in strongly correlated electron systems.
 \end{abstract}

\maketitle

Correlated electrons have exhibited many interesting phenomena that deviate from the Fermi liquid theory and the theory of phase transition.  Taking the high transition temperature superconductors in cuprates as an example~\cite{bednorz1986}, the pseudogap formation~\cite{shen2003}, charge-density-wave fluctuations observed in the scanning tunneling microscopes and resonant soft x-ray scattering experiments ~\cite{keimer2012, davis2014, davis2016a, davis2016b}, and ferromagnetic time-reversal-symmetry-breaking fluctuations~\cite{kapitulnik2008}, occur simultaneously wide in the phase diagram.  The onset of the time-reversal-symmetry-breaking fluctuation coincides with formation temperature of the pseudogap~\cite{kapitulnik2008}.  The charge-density-wave fluctuation resides well in the pseudogap phase~\cite{taillefer2016}.  Those fluctuations have one common property.  Namely, there are no signatures of phase transition as they occur. Their origins are mysterious.  In this paper, we will demonstrate that charge-density-wave fluctuation and time-reversal-symmetry-breaking fluctuation are actually universal, if the correlated electrons have the pseudogap phase.

Recently, one of us (CHC) proposed the theory of the pseudogap formation~\cite{chern2014}.  The electrons \emph{weakly} interacting with the U(1) gauge field, originated from the spin Berry's phase~\cite{nagaosa2014}, open a gap-like structure, when the gauge field acquires the mass.  The mass acquisition of the gauge field is due to the \emph{strong coupling} with the anti-ferromagnetic \emph{fluctuation}, a remnant of the anti-ferromagnetism as the system is doped.  The basic assumption of this theory is that the spin anisotropy is a relevant perturbation, so that the anti-ferromagnetic fluctuation can be described by a phase field, $\phi(\vec{x},t)=\frac{1}{q}e^{i\sigma(\vec{x},t)}$, where $q$ is the coupling between the gauge field and the anti-ferromagnetic fluctuation. We emphasize that the anti-ferromagnetic fluctuation does not couple to the elections directly.  In two dimensions, the Kosterlitz-Thouless (KT) transition takes place for the phase field at finite temperature.  Then, the anti-ferromagnetic fluctuation is absorbed by the gauge transformation and becomes the \emph{longitudinal} component of the gauge field.  Because the gauge field acquires the mass, the interaction between electrons becomes short-ranged.    Due to the nature of the KT transition, there are no conventional signatures of phase transitions.  Translational symmetry and the time reversal symmetry are well preserved.

At the first glance, it looks contradictory that the phase, preserving both the translational and time reversal symmetries, hosts the charge-density-wave fluctuation and the time-reversal-symmetry-breaking fluctuation.  We will show later that they are fluctuations and not the ordering states.  Electronic interaction mediated by the gauge field infers that electrons exchange virtual particles of the pure imaginary wave vectors.  Nonetheless, due to the quantum fluctuation, gauge field can be excited in the propagation modes of the real wave vector.  The charge-density-wave fluctuation is the direct consequence of the propagating \emph{gauge-electric field} contributed from the \emph{longitudinal} mode.  On the other hand, the ferromagnetic time-reversal symmetry-breaking fluctuation originates from the propagating \emph{gauge-magnetic field} of the transverse modes.

This paper is organized as the following.  We will discuss the effective interaction between electrons by integrating out the electronic degree of freedom.  The propagation modes of the gauge field can be obtained by solving the classical equations of motion.  Then, we consider the classical motion of the electrons in the presence of the propagating gauge field.  We will apply the current scheme to the high-$T_c$ superconductors.  The presence of the anti-ferromagnetic fluctuation and the emergence of the gauge interaction baptize the quantum correlation.  Once it is considered carefully, many of pseudogap phenomenology can be realized.  The pseudogap formation, the charge-density-wave fluctuation, and the time-reversal-symmetry-breaking fluctuation do not have the relation of causality.  They are all the consequent phenomena of the correlation.

\begin{figure}[htb]
\includegraphics[width=0.2\textwidth]{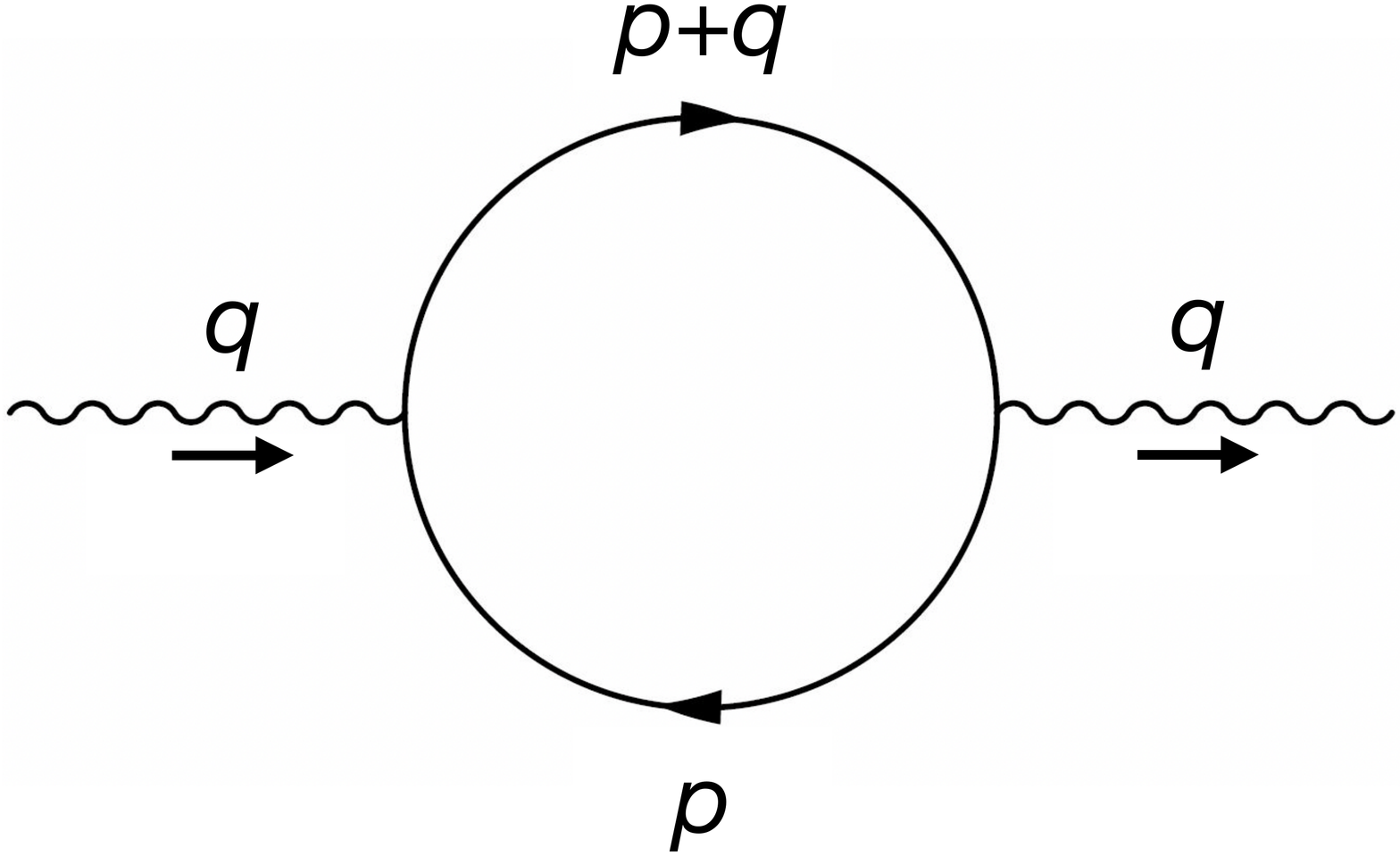}
\includegraphics[width=0.2\textwidth]{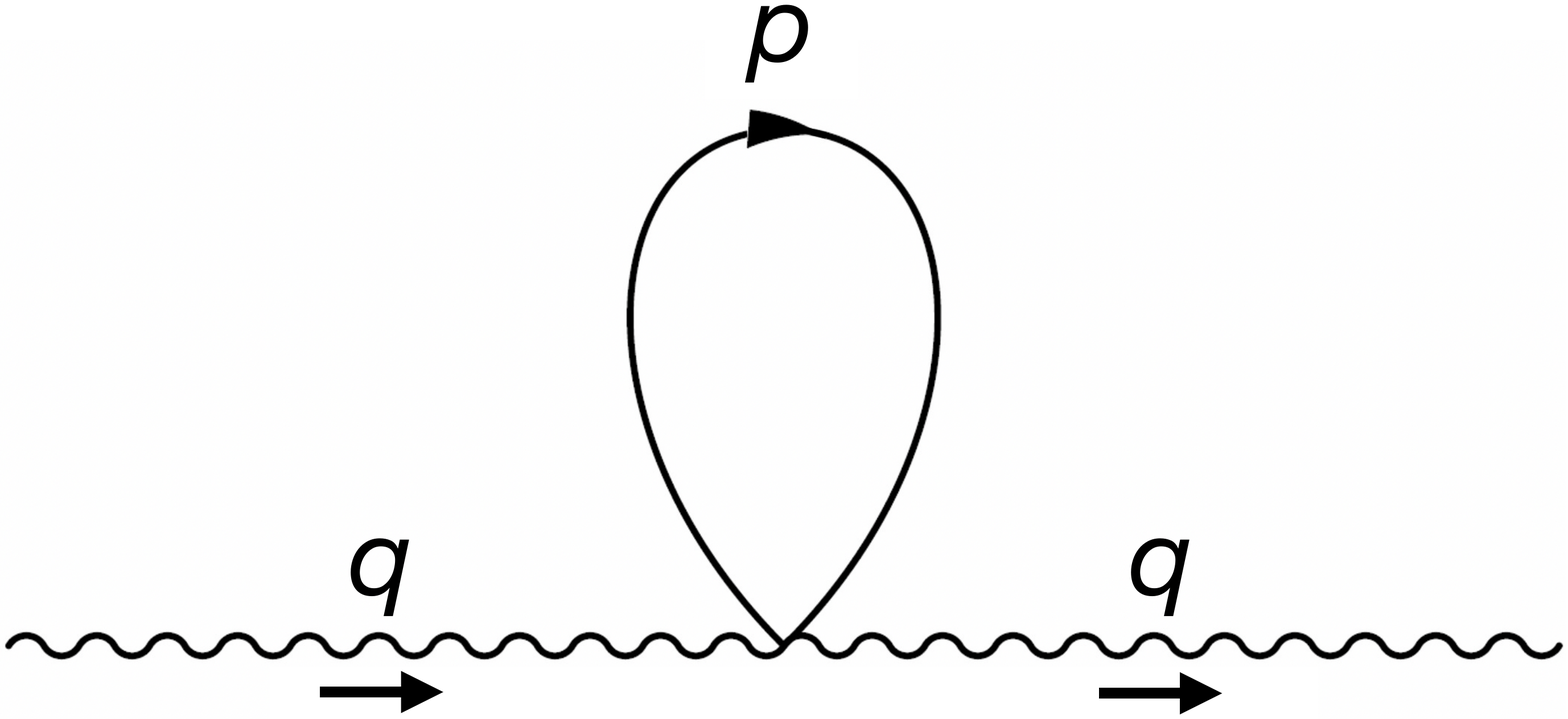}
\caption{Feynman diagrams to compute the effective Lagrangian of the gauge field, that is proportional to $f_{\mu\nu}f_{\mu\nu}$ in the long wavelength limit.} \label{Fig:FF}
\end{figure}

As the cuprates are doped, the anti-ferromagnetic ordering ceases, the pseudogap phase is developed, and the gapless states are generated in the nodal directions.  It turns out that pseudogap structure is anisotropic in the momentum space, which we believe that it is the sum of the two causes: one mechanism to open an isotropic gap~\cite{chern2014} and the another mechanism to introduce  the nodal quasiparticles~\cite{Lee2018}.  In this paper, we do not explain the phenomena associated with the nodal quasiparticles.  We focus on the consequences that relate to the pseudogap. 
Let us consider the following Lagrangian density~\cite{chern2014}
\begin{eqnarray}
\mathscr{L}\! &=&\! \mathscr{L}_{\psi}+\mathscr{L}_{a}+\mathscr{L}_{\phi}, \nonumber\\
&\mathscr{L}_{\psi}&=\!\!\sum_{\alpha}\psi^\dag_{\alpha}(i\partial_0)\psi_\alpha\!-\!\frac{1}{2m}[(-\frac{\vec{\nabla}}{i}\!-\!g\vec{a})\psi^\dag_\alpha][(\frac{\vec{\nabla}}{i}\!-\!g\vec{a})\psi_\alpha]\nonumber\\&&-ga_0\psi^\dag_\alpha\psi_\alpha, \nonumber\\
&\mathscr{L}_{a}&=\!-\!\frac{1}{4}f_{\mu\nu}f_{\mu\nu},\nonumber \\ 
&\mathscr{L}_{\phi}&=\!\!\frac{1}{2}M_0^2(D_0 \phi)^\dag\!(D_0 \phi)\!-\!\frac{1}{2}M_1^2(D_i \phi)^\dag\!(D_i \phi),\label{fulllagrangian}
\end{eqnarray}
where $\psi_\alpha$ is the electron variable with the spin index $\alpha$, ($a_0$, $\vec{a}$) is the gauge field, $g$ is the coupling of the electrons to the gauge field, and $M_0$ and $M_1$ are the mass parameters.    In Eq.(\ref{fulllagrangian}), we adopted the natural unit, where 
$\hbar$ and the speed of light $c$ are set to be 1.  It is equivalent to roughly set 197 eV$\cdot$ nm = 1, which indicates that the mass of the gauge field defines the length scale.  In cuprates, the wavelength of the charge density wave appears to be the only length scale, which implies $M_0=M_1$.  Considering together the pseudogap magnitude, about 40 meV~\cite{shen2014}, the dimensionless gauge coupling $\frac{g^2}{2m}$ can be computed $\sim 1.5\times10^{-3}$~\cite{chern2014}.  The weak-coupling nature allows us to compute the effective Lagrangian of the gauge field and the $\phi$ field perturbatively.  Integrating out the electronic degrees of freedom, the diagrams that renormalize the gauge coupling are given in Fig.~\ref{Fig:FF}.    Using the Green's function of electrons for the insulators~\cite{fujimori1997, fujimori2006}, those diagrams are zero.  Namely, the gauge coupling is not renormalized by the electrons. 

This result implies that the pseudogap magnitude and the onset temperature are independent of the external magnetic field~\cite{gorny1999, zheng1999, gorny2001, mitrivic2002, zheng2012}.  It is because the external magnetic field couples only to the electrons, and they have no contribution to renormalize the gauge coupling and the mass of the gauge field.

Having integrated out the electron degrees of freedom, the classical equations of motion of the gauge field and the $\phi = \frac{1}{q}e^{i\sigma}$ field can be derived.
\begin{eqnarray}
&&\vec{\nabla}\cdot(\vec{\nabla}a_0+\partial_t\vec{a}) = M_0^2(\frac{1}{q}\partial_t \sigma +a_0) \nonumber \\
&&\vec{\nabla}\times(\vec{\nabla}\times \vec{a}) = M_1^2(\frac{1}{q}\vec{\nabla}\sigma - \vec{a})-\partial_t(\vec{\nabla}a_0+\partial_t\vec{a}) \nonumber \\
&&M_0^2\partial_t(\frac{1}{q}\partial_t\sigma + a_0) = M_1^2\vec{\nabla}\cdot(\frac{1}{q}\vec{\nabla}\sigma -\vec{a}). \label{eom}
\end{eqnarray} 
The Hamiltonian density can be also computed.
\begin{eqnarray}
\mathscr{H} = \!&\frac{1}{2}&\!(E^2+B^2) + (\frac{1}{q}\partial_t\sigma +a_0)\vec{\nabla}\cdot(\vec{\nabla}a_0+\partial_t\vec{a})\nonumber \\\!\!&-&\!\!\frac{M_0^2}{2}(\frac{1}{q}\partial_t\sigma+a_0)^2+\frac{M_1^2}{2}(\frac{1}{q}\vec{\nabla}\sigma-\vec{a})^2, \label{hami}
\end{eqnarray}
where $\vec{E}=-\vec{\nabla}a_0-\partial_t \vec{a}$ is the gauge-electric field and $\vec{B}=\vec{\nabla}\times \vec{a}$ is the gauge-magnetic field.  Solving Eq.(\ref{eom}) in the pseudogap phase, where the expectation value of $\sigma$ takes 0, we obtain 
\begin{eqnarray}
&&(\partial^2_t -\frac{M_1^2}{M_0^2}\nabla^2+M_1^2)a_0 =0 \nonumber \\
&&(\partial^2_t-\nabla^2 +M_1^2)\vec{a}=(\frac{M_0^2}{M^2_1}-1)\partial_t(\vec{\nabla} a_0) \label{wavequ}
\end{eqnarray}
There are two solutions in Eq.(\ref{wavequ}).  The longitudinal mode has the dispersion relation $\omega_L^2 = \frac{M_1^2}{M_0^2}k_L^2+M_1^2$, and the transverse mode has the dispersion relation $\omega_T^2 = k_T^2+M_1^2$.  

In the high temperature phase, the $\phi(\vec{x},t)$ field is fluctuating.  The gauge field is massless containing only the transverse mode.  In the pseudogap phase, the $\phi(\vec{x},t)$ field picks up a quasi-long-ranged order through the Kosterlitz-Thouless transition and becomes the longitudinal mode of the gauge field via the gauge transformation~\cite{chern2014}.  Interestingly, the longitudinal mode has only $\vec{E}$ field and no $\vec{B}$ field.  Excited by the quantum fluctuation, the $\vec{E}$ field gives the non-trivial dynamics to the electrons.  Without losing generality, we consider the standing-wave solution and take the $x$ direction as the longitudinal direction, $a_0 = A_0 e^{ik_Lx}\cos(\omega_L t)$ and $\vec{a}=\frac{-ik_L\omega_L}{\omega_L^2-M^2_1}A_0e^{ik_Lx}\sin(\omega_L t)\hat{x}$, where $A_0$ is the strength of the quantum fluctuation, and its magnitude will be determined shortly.  The energy density of the longitudinal mode can be computed $\mathscr{E}_L = \frac{A^2_0}{4}(\frac{M_0^4}{k_L^2}+M_0^2)$.  Likewise, the energy density of the transverse mode can be computed $\mathscr{E}_T = \frac{A^2_1}{4}(k^2_T+M_1^2)$, if $\vec{a} = A_1e^{ik_Tx}\cos(\omega_T t)\hat{y}$, where $A_1$ is the strength of the quantum fluctuation.

Apparently, the longitudinal mode and the transverse mode have very different characters.  From their energy density, the longitudinal mode favors a big $k_L$, and the transverse mode favors a long wavelength $k_T$.  Therefore, the $\vec{E}$ field modulation of the longitudinal mode must be in the lattice scale, and the $\vec{B}$ field of the transverse mode favors the uniform distribution.  As we will see later, the former is the driving force of the charge-density-wave fluctuation.  Driven by the longitudinal mode, the electrons acquire the kinetic energy to form the orbital magnetic moment in the presence of the uniform $\vec{B}$ field of the transverse mode, resulting in the ferromagnetic time-reversal-symmetry-breaking fluctuation in the polar Kerr rotation experiments.

Let us consider the classical dynamics of the electrons in the presence of $\vec{E}$ field of the longitudinal mode which is given by $\vec{E}=i\frac{A_0M^2_0}{k_L}e^{ik_Lx}\cos(\omega_L t)\hat{x}$.
Taking the one that the origin is the node, It causes the acceleration 
\begin{eqnarray}
\ddot{\vec{x}}(t) = -\frac{gA_0M_0^2}{mk_L}\sin(k_Lx)\cos(\omega_Lt)\hat{x}. \label{acc}
\end{eqnarray}
We solve Eq.(\ref{acc}) numerically.  The initial conditions take the state of uniform density with zero initial velocity.  The values of the parameters are taken from the experiments.  Namely, the wavelength is about 4 lattice constants that is around 1.6 nm.  In the natural unit, $k_L = \frac{2\pi}{1.6 \text{nm}}$ $\doteq$ 773 eV, $M_0 = \frac{1}{1.6 \text{nm}} \doteq$ 123 eV, $g$ = 39.13 $\sqrt{\text{eV}}$, and the electron mass $m$ = 0.5 MeV.  The $A_0$ is strength of the quantum fluctuation of $\vec{E}$ field of the energy $\frac{\mathscr{E}_L}{M_0^2}=\frac{A^2_0}{4}(\frac{M_0^2}{k_L^2}+1)=\frac{1}{4}(\frac{1}{4\pi^2}+1)A^2_0$, which should be an experimentally determined parameter.  Due to the energy conservation, it should correspond to the onset temperature of the charge-density-wave fluctuation, namely $\frac{1}{4}(\frac{1}{4\pi^2}+1)A^2_0 \sim k_BT_{\text{CDW}}$.   For $T_{CDW}=100 K$, $A_0 = 0.183 \sqrt{\text{eV}}$.  It is a quantum fluctuation in the sense that the energy of the $\vec{E}$ field is much smaller than the energy scale of $M_0$.  Taking the length in the unit of the wavelength $\tilde{x}=x/\lambda_{\text{CDW}}$ and the time in the unit of the period $\tilde{t}=M_0t$, the dimensionless acceleration $\ddot{\vec{X}}(\tilde{t}) = \frac{\ddot{\vec{x}}(t)}{k_L} = -\frac{gA_0M_0^2}{mk_L^2}\sin(2\pi\tilde{x})\cos(\sqrt{1+4\pi^2}\tilde{t})\hat{x}$.  The magnitude of the dimensionless acceleration $\frac{gA_0M_0^2}{mk_L^2}$ is roughly $3.6\times 10^{-7}$ in cuprates. 

\begin{figure}[htb]
\includegraphics[width=0.4\textwidth]{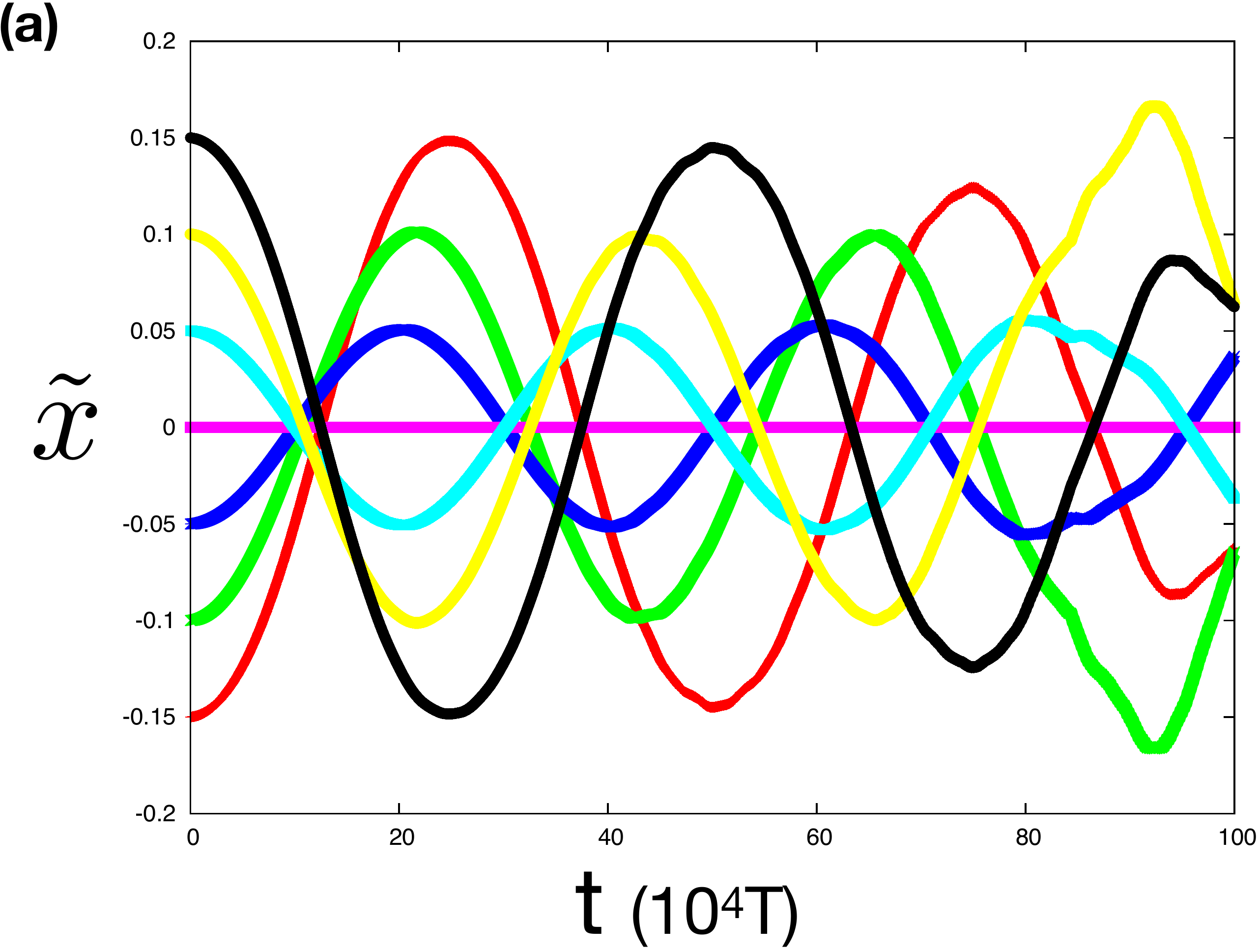}
\includegraphics[width=0.4\textwidth]{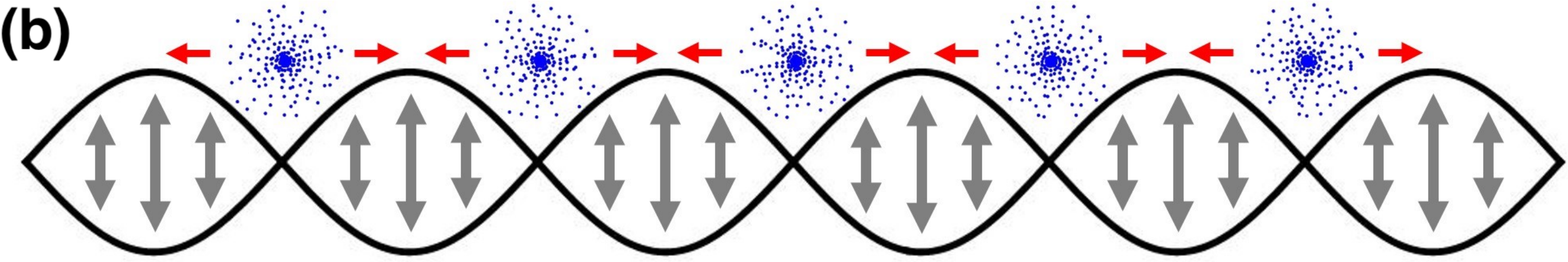}
\caption{(Color online) (a) Eq.(\ref{acc}) is solved numerically for several initial positions.  $\tilde{x}$ is dimensionless in the unit of the wavelength.  The time $t$ is in the unit of the period.  In this figure, 3.6 $\times 10^{-6}$ of the acceleration magnitude is used for the better data illustration.  Physics is the same for 3.6 $\times 10^{-7}$ acceleration, though uncontrollable numerical error hinders to see more oscillations around the nodes.  In the main text, all results are based on the calculations using the 3.6 $\times 10^{-7}$ acceleration.  (b) The cartoon to illustrate the electron motions.  Blue dots represent the electron.  Black curves represent the standing wave of the $\vec{E}$ field.}\label{Fig:XofT}
\end{figure}

\begin{figure}[htb]
\includegraphics[width=0.4\textwidth]{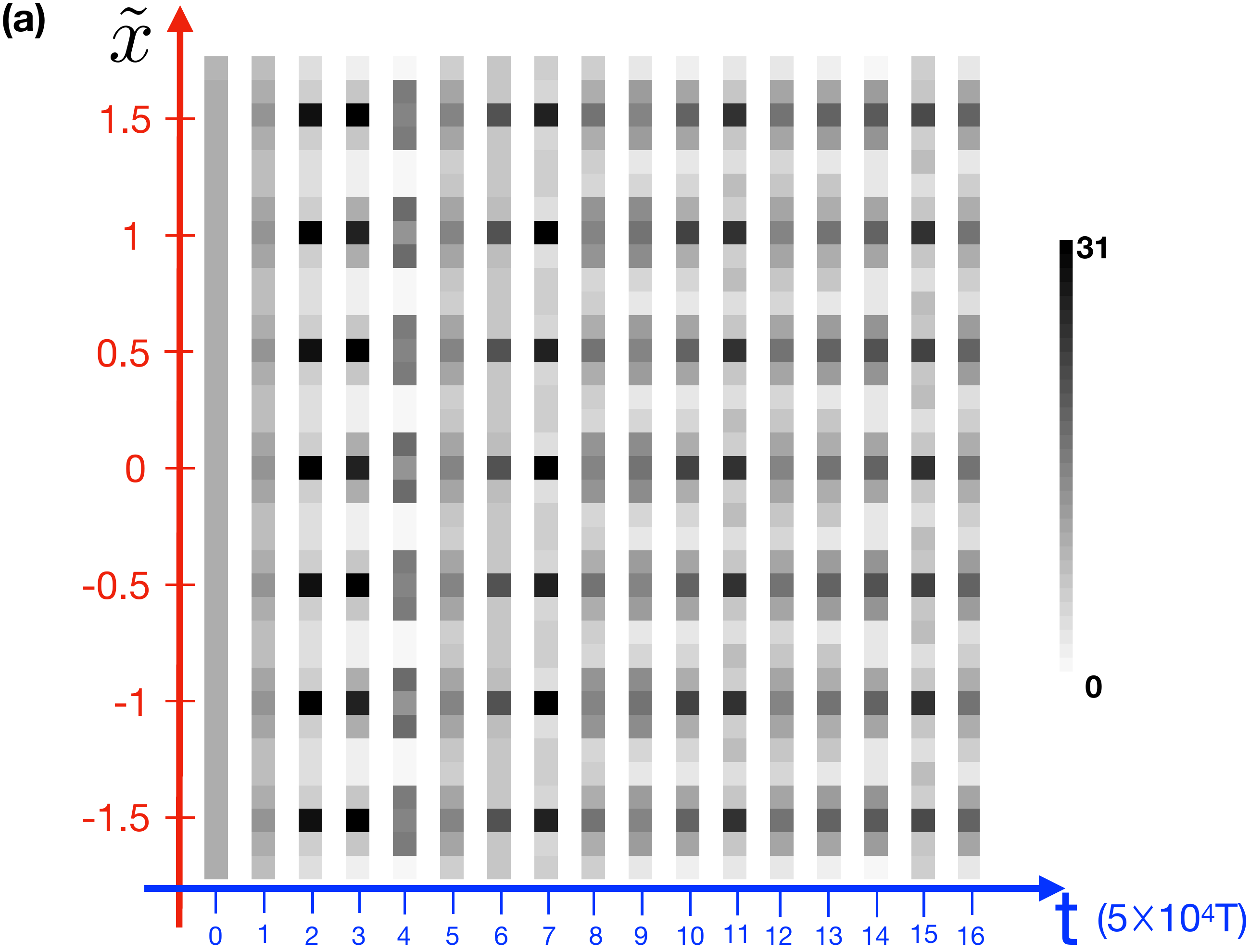}
\includegraphics[width=0.4\textwidth]{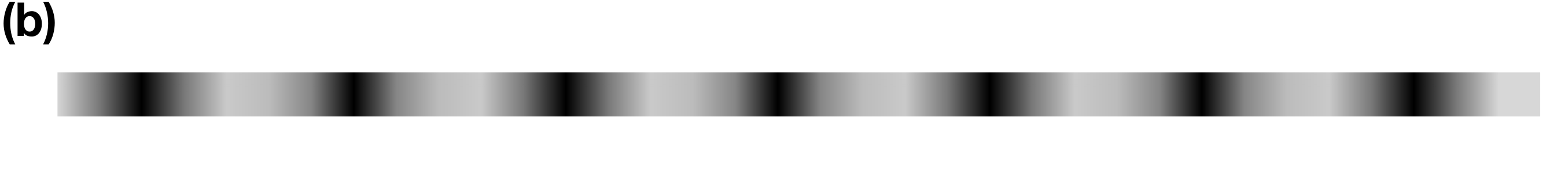}
\caption{(Color online) (a) The snap shots of the density patterns taken at every $5\times 10^4$ oscillations which is equivalent to $2.67\times 10^{-13}$ seconds. (data of $3.6\times 10^{-6}$ acceleration) (b) The time average of the electron density over $4.2\times 10^{-12}$ seconds.}\label{Fig:density}
\end{figure}

In Fig.(\ref{Fig:XofT}a), we show the $\tilde{x}(t)$ for several initial positions.  The standing wave of the $\vec{E}$ field confines the electrons to move around the nodes.  The period $T$ of the standing wave of the $\vec{E}$ field is about $5.34\times10^{-18}$ seconds.  Since the acceleration is weak and the oscillation of the $\vec{E}$ field is fast, electrons shudder tiny distance in every oscillation of the $\vec{E}$ field.  It turns out that they take about $10^6$ oscillations to reach to the nodes, that is equivalent to the order of $10^{-11}$ seconds.  Different initial positions take different period to move around the nodes, leading to a time-dependent pattern of the electron density modulation.  We can take snap shots of the density wave, and the results are given in Fig.(\ref{Fig:density}a).  In our calculations, 349 electrons distribute uniformly over 3.5 $\lambda_{\text{CDW}}$ at $t = 0$. Here we use high electron density just for the better data visualization.   As all electrons oscillate around the nodes, it generates a modulation of electron density in the real space.  In Fig.(\ref{Fig:density}b), the time average of the density patterns is computed. 

We emphasize that this effect is independent of the pseudogap formation.  The charge-density-wave fluctuation needs the propagating longitudinal mode of the real wave vectors, but the pseudogap formation takes the virtual longitudinal mode of the imaginary wave vectors.  It is very similar to the electrodynamics, where there are both virtual photons to mediate the electromagnetic interaction and the real electromagnetic waves.  Consequently, it is possible that a system has the pseudogap and no charge density wave is observed if the quantum fluctuation is zero.  Therefore, the onset temperature of the charge-density-wave fluctuation is not necessarily the same as the one of the pseudogap formation.  Especially, in the heavily underdoped systems, the low electron mobility additionally hampers the development of the modulation.  Nevertheless, it is also possible that the modulation is too weak to be detected.

Let us also comment on the wavelength of charge density wave, which is measured few lattice constants in experiments.  As mentioned earlier, the $\vec{E}$ field of the longitudinal mode favors a big $k_L$.  However, the theory allows different wavelengths for different systems.  Since $A_0$ is $k_L$ dependent, systems determine the most energetically favorable channel of $k_L$, which can be commensurate or incommensurate.  We do not believe that it is predictable.  

Furthermore, the nodes of the standing wave locate at the phases 0 and $\pi$, where the anti-nodes of the charge density wave locate.  For the systems of multi-electrons in a unit cell, the phases of the anti-nodes for the different electrons in the same node of the $\vec{E}$ field can be differed by 0 or $\pi$.  For example, if there are two electrons in the unit cell, the phases of these two density waves can be the same or have $\pi$ difference.  In different domains, the phase difference can be different, as well.  

In conclusion, the occurrence of the fluctuating $\vec{E}$ field of the longitudinal mode provides the external driving force for  the modulation formation.  It is very similar to the water ripples blown by the wind.  In fact, using a plane-wave solution of the $\vec{E}$ field, we obtain a propagating modulation of density, which we believe that it also happens in real systems.  Unlike an ordering state due to the electron-electron or the electron-phonon interaction, this mechanism does not need a phase transition.

Taking the curl on the 2$^{\text{nd}}$ equation in Eq.(\ref{wavequ}), we have $(\partial^2_t-\nabla^2 +M_1^2)\vec{B}=0$.  The uniform solution then satisfies the equation $(\partial^2_t +M_1^2)\vec{B}=0$, and $\vec{B}=B_0\sin(M_1t+\delta)\hat{z}$, where $\delta$ are the initial phases and $\vec{B}$ only has the $z$ component in two dimensional systems.  The energy density is given by $\mathscr{E}_{B}=\frac{B^2_0}{4}[1+2M_1^2(x+x_0)^2]$, where $x_0$ is an arbitrary reference point and the magnetic energy and the electric induction energy is included.  Again, roughly taking the onset temperature $\sim$ 100 K, the strength of the quantum fluctuation $B_0$ is 21.14 (eV)$^{\frac{3}{2}}$, that corresponds to 14.12 Tesla.  In general, $\delta$ are different in different domains.  In order to reveal the ferromagnetic signal, the quantum fluctuations has to be in phase.  In the polar Kerr rotation experiments, a magnetic field of $4$ Tesla is applied to achieve this before the measurement.

The ferromagnetic time-reversal-symmetry-breaking fluctuation is due to the orbital motion of the electron in the presence of the quantum fluctuation of the uniform $B$ field.  However, the magnetic moments by the circular motion of electrons under the $\vec{B}$ field simply cancel out, since it oscillates rapidly in the frequency $1/M_1 \sim 2\times 10^{17}$ Hz.  In the presence of the $\vec{E}$ field of the longitudinal mode, electrons move around the nodes collectively.  In the real systems, the $\vec{E}$ field can be excited in both $x$ and $y$ directions, and the nodes distribute at $\frac{1}{2}\lambda_{\text{CDW}}(m,n)$, where $m$ and $n$ are integers.  Then, electrons move around the nodes in the "radial" directions.  Now, in the presence of the $B$ field fluctuation, Lorentz force provides the centrifugal force and electrons \emph{rotate} around the nodes.  The rotational direction  depends on the initial phase of the magnetic field.  Although the $B$ field is rapidly oscillating, the initial phase determines the tangential velocity in the begining and thus determines whether the orbital motion is clockwise or counter-clockwise.  The period is determined by the velocity of the collective motion, that is $\sim 2\times 10^{-11}$ seconds.  The radius of the loop is about one quarter of the wavelength of the density modulation, that is $\sim 0.4$ nm.  Consequently, the magnetic moment is in the order of $10^{-4}\mu_{\text{B}}$, where $\mu_{\text{B}}$ is the Bohr magneton. 

In the experiments, the training field sometimes results in the totally opposite magnetic moments.  For example, in Ref.\cite{kapitulnik2008}, the training field of 3 Tesla has different results from the 4 Tesla ones.  It is due to the initial phase $\delta$.  For example, the tangential velocity for $\delta=\pi$ takes the opposite direction to the one for $\delta=0$.  The magnetic moment then reverses.  We believe that even the training field at 4 Tesla could have both results.  The training-field-dependent physics is the best evidence for the existence of the fluctuating magnetic field.  Furthermore, this mechanism dictates that the time-reversal-symmetry fluctuation has the close relation with the modulation formation.  We predict that their onset temperature should be correlated.

We have provided the clear physical pictures and quantitatively described the origins of the charge-density-wave fluctuation and the ferromagnetic time-reversal-symmetry fluctuation.  Those effects are the strong evidences of the existence of the gauge interaction, that also leads to the pseudogap formation.  Once the gauge interaction is properly considered, the phenomena in the correlated electron systems are as traditional as Physics 101.  We believe that they are universal, if the correlated electron systems have pseudogap and finite quantum fluctuations.  The pseudogap formation, the charge-density-wave fluctuation, and the time-reversal-symmetry-breaking fluctuation are the basic notions in the strongly correlated electron systems.


CHC acknowledges the stimulating discussion with Naoto Nagaosa.  He is particularly in debt to Dung-Hai Lee for the discussions on the pseudogap phenomenology.  This work is supported by MOST 106-2112-M-002-007-MY3 of Taiwan.

\begin{thebibliography}{19}%
\makeatletter
\providecommand \@ifxundefined [1]{%
 \@ifx{#1\undefined}
}%
\providecommand \@ifnum [1]{%
 \ifnum #1\expandafter \@firstoftwo
 \else \expandafter \@secondoftwo
 \fi
}%
\providecommand \@ifx [1]{%
 \ifx #1\expandafter \@firstoftwo
 \else \expandafter \@secondoftwo
 \fi
}%
\providecommand \natexlab [1]{#1}%
\providecommand \enquote  [1]{``#1''}%
\providecommand \bibnamefont  [1]{#1}%
\providecommand \bibfnamefont [1]{#1}%
\providecommand \citenamefont [1]{#1}%
\providecommand \href@noop [0]{\@secondoftwo}%
\providecommand \href [0]{\begingroup \@sanitize@url \@href}%
\providecommand \@href[1]{\@@startlink{#1}\@@href}%
\providecommand \@@href[1]{\endgroup#1\@@endlink}%
\providecommand \@sanitize@url [0]{\catcode `\\12\catcode `\$12\catcode
  `\&12\catcode `\#12\catcode `\^12\catcode `\_12\catcode `\%12\relax}%
\providecommand \@@startlink[1]{}%
\providecommand \@@endlink[0]{}%
\providecommand \url  [0]{\begingroup\@sanitize@url \@url }%
\providecommand \@url [1]{\endgroup\@href {#1}{\urlprefix }}%
\providecommand \urlprefix  [0]{URL }%
\providecommand \Eprint [0]{\href }%
\providecommand \doibase [0]{http://dx.doi.org/}%
\providecommand \selectlanguage [0]{\@gobble}%
\providecommand \bibinfo  [0]{\@secondoftwo}%
\providecommand \bibfield  [0]{\@secondoftwo}%
\providecommand \translation [1]{[#1]}%
\providecommand \BibitemOpen [0]{}%
\providecommand \bibitemStop [0]{}%
\providecommand \bibitemNoStop [0]{.\EOS\space}%
\providecommand \EOS [0]{\spacefactor3000\relax}%
\providecommand \BibitemShut  [1]{\csname bibitem#1\endcsname}%
\let\auto@bib@innerbib\@empty
\bibitem [{\citenamefont {Bednorz}\ and\ \citenamefont
  {M\"uller}(1986)}]{bednorz1986}%
  \BibitemOpen
  \bibfield  {author} {\bibinfo {author} {\bibfnamefont {J.~G.}\ \bibnamefont
  {Bednorz}}\ and\ \bibinfo {author} {\bibfnamefont {K.~A.}\ \bibnamefont
  {M\"uller}},\ }\href@noop {} {\bibfield  {journal} {\bibinfo  {journal} {Z.
  Phys. B: Condens. Matter}\ }\textbf {\bibinfo {volume} {64}},\ \bibinfo
  {pages} {189} (\bibinfo {year} {1986})}\BibitemShut {NoStop}%
\bibitem [{\citenamefont {Damascelli}\ \emph {et~al.}(2003)\citenamefont
  {Damascelli}, \citenamefont {Hussain},\ and\ \citenamefont
  {Shen}}]{shen2003}%
  \BibitemOpen
  \bibfield  {author} {\bibinfo {author} {\bibfnamefont {A.}~\bibnamefont
  {Damascelli}}, \bibinfo {author} {\bibfnamefont {Z.}~\bibnamefont {Hussain}},
  \ and\ \bibinfo {author} {\bibfnamefont {Z.-X.}\ \bibnamefont {Shen}},\
  }\href@noop {} {\bibfield  {journal} {\bibinfo  {journal} {Rev. Mod. Phys.}\
  }\textbf {\bibinfo {volume} {75}},\ \bibinfo {pages} {473} (\bibinfo {year}
  {2003})}\BibitemShut {NoStop}%
\bibitem [{\citenamefont {Ghiringhelli}\ \emph {et~al.}(2012)\citenamefont
  {Ghiringhelli}, \citenamefont {Tacon}, \citenamefont {Minola}, \citenamefont
  {Blanco-Canosa}, \citenamefont {Mazzoli}, \citenamefont {Brookes},
  \citenamefont {Luca}, \citenamefont {Frano}, \citenamefont {Hawthorn},
  \citenamefont {He}, \citenamefont {Loew}, \citenamefont {Sala}, \citenamefont
  {Peets}, \citenamefont {Salluzzo}, \citenamefont {Schierle}, \citenamefont
  {Sutarto}, \citenamefont {Sawatzky}, \citenamefont {Weschke}, \citenamefont
  {Keimer},\ and\ \citenamefont {Braicovich}}]{keimer2012}%
  \BibitemOpen
  \bibfield  {author} {\bibinfo {author} {\bibfnamefont {G.}~\bibnamefont
  {Ghiringhelli}}, \bibinfo {author} {\bibfnamefont {M.~L.}\ \bibnamefont
  {Tacon}}, \bibinfo {author} {\bibfnamefont {M.}~\bibnamefont {Minola}},
  \bibinfo {author} {\bibfnamefont {S.}~\bibnamefont {Blanco-Canosa}}, \bibinfo
  {author} {\bibfnamefont {C.}~\bibnamefont {Mazzoli}}, \bibinfo {author}
  {\bibfnamefont {N.~B.}\ \bibnamefont {Brookes}}, \bibinfo {author}
  {\bibfnamefont {G.~M.~D.}\ \bibnamefont {Luca}}, \bibinfo {author}
  {\bibfnamefont {A.}~\bibnamefont {Frano}}, \bibinfo {author} {\bibfnamefont
  {D.~G.}\ \bibnamefont {Hawthorn}}, \bibinfo {author} {\bibfnamefont
  {F.}~\bibnamefont {He}}, \bibinfo {author} {\bibfnamefont {T.}~\bibnamefont
  {Loew}}, \bibinfo {author} {\bibfnamefont {M.~M.}\ \bibnamefont {Sala}},
  \bibinfo {author} {\bibfnamefont {D.~C.}\ \bibnamefont {Peets}}, \bibinfo
  {author} {\bibfnamefont {M.}~\bibnamefont {Salluzzo}}, \bibinfo {author}
  {\bibfnamefont {E.}~\bibnamefont {Schierle}}, \bibinfo {author}
  {\bibfnamefont {R.}~\bibnamefont {Sutarto}}, \bibinfo {author} {\bibfnamefont
  {G.~A.}\ \bibnamefont {Sawatzky}}, \bibinfo {author} {\bibfnamefont
  {E.}~\bibnamefont {Weschke}}, \bibinfo {author} {\bibfnamefont
  {B.}~\bibnamefont {Keimer}}, \ and\ \bibinfo {author} {\bibfnamefont
  {L.}~\bibnamefont {Braicovich}},\ }\href@noop {} {\bibfield  {journal}
  {\bibinfo  {journal} {Science}\ }\textbf {\bibinfo {volume} {337}},\ \bibinfo
  {pages} {821} (\bibinfo {year} {2012})}\BibitemShut {NoStop}%
\bibitem [{\citenamefont {Fujita}\ \emph {et~al.}(2014)\citenamefont {Fujita},
  \citenamefont {Hamidian}, \citenamefont {Edkins}, \citenamefont {Kim},
  \citenamefont {Kohsaka}, \citenamefont {Azuma}, \citenamefont {Takano},
  \citenamefont {Takagi}, \citenamefont {Eisaki}, \citenamefont {Uchida},
  \citenamefont {Allais}, \citenamefont {Lawler}, \citenamefont {Kim},
  \citenamefont {Sachdev},\ and\ \citenamefont {Davis}}]{davis2014}%
  \BibitemOpen
  \bibfield  {author} {\bibinfo {author} {\bibfnamefont {K.}~\bibnamefont
  {Fujita}}, \bibinfo {author} {\bibfnamefont {M.~H.}\ \bibnamefont
  {Hamidian}}, \bibinfo {author} {\bibfnamefont {S.~D.}\ \bibnamefont
  {Edkins}}, \bibinfo {author} {\bibfnamefont {C.~K.}\ \bibnamefont {Kim}},
  \bibinfo {author} {\bibfnamefont {Y.}~\bibnamefont {Kohsaka}}, \bibinfo
  {author} {\bibfnamefont {M.}~\bibnamefont {Azuma}}, \bibinfo {author}
  {\bibfnamefont {M.}~\bibnamefont {Takano}}, \bibinfo {author} {\bibfnamefont
  {H.}~\bibnamefont {Takagi}}, \bibinfo {author} {\bibfnamefont
  {H.}~\bibnamefont {Eisaki}}, \bibinfo {author} {\bibfnamefont
  {S.}~\bibnamefont {Uchida}}, \bibinfo {author} {\bibfnamefont
  {A.}~\bibnamefont {Allais}}, \bibinfo {author} {\bibfnamefont {M.~J.}\
  \bibnamefont {Lawler}}, \bibinfo {author} {\bibfnamefont {E.~A.}\
  \bibnamefont {Kim}}, \bibinfo {author} {\bibfnamefont {S.}~\bibnamefont
  {Sachdev}}, \ and\ \bibinfo {author} {\bibfnamefont {J.~C.~S.}\ \bibnamefont
  {Davis}},\ }\href@noop {} {\bibfield  {journal} {\bibinfo  {journal} {Proc.
  Nat'l. Acad. Sci.}\ }\textbf {\bibinfo {volume} {111}},\ \bibinfo {pages}
  {E3026} (\bibinfo {year} {2014})}\BibitemShut {NoStop}%
\bibitem [{\citenamefont {Hamidian}\ \emph {et~al.}(2016)\citenamefont
  {Hamidian}, \citenamefont {Edkins}, \citenamefont {Kim}, \citenamefont
  {Davis}, \citenamefont {Mackenzie}, \citenamefont {Eisaki}, \citenamefont
  {Uchida}, \citenamefont {Lawler}, \citenamefont {Kim}, \citenamefont
  {Sachdev},\ and\ \citenamefont {Fujita}}]{davis2016a}%
  \BibitemOpen
  \bibfield  {author} {\bibinfo {author} {\bibfnamefont {M.~H.}\ \bibnamefont
  {Hamidian}}, \bibinfo {author} {\bibfnamefont {S.~D.}\ \bibnamefont
  {Edkins}}, \bibinfo {author} {\bibfnamefont {C.-K.}\ \bibnamefont {Kim}},
  \bibinfo {author} {\bibfnamefont {J.~C.}\ \bibnamefont {Davis}}, \bibinfo
  {author} {\bibfnamefont {A.~P.}\ \bibnamefont {Mackenzie}}, \bibinfo {author}
  {\bibfnamefont {H.}~\bibnamefont {Eisaki}}, \bibinfo {author} {\bibfnamefont
  {S.}~\bibnamefont {Uchida}}, \bibinfo {author} {\bibfnamefont {M.~J.}\
  \bibnamefont {Lawler}}, \bibinfo {author} {\bibfnamefont {E.-A.}\
  \bibnamefont {Kim}}, \bibinfo {author} {\bibfnamefont {S.}~\bibnamefont
  {Sachdev}}, \ and\ \bibinfo {author} {\bibfnamefont {K.}~\bibnamefont
  {Fujita}},\ }\href@noop {} {\bibfield  {journal} {\bibinfo  {journal} {Nature
  Physics}\ }\textbf {\bibinfo {volume} {12}},\ \bibinfo {pages} {150}
  (\bibinfo {year} {2016})}\BibitemShut {NoStop}%
\bibitem [{\citenamefont {Mesaros}\ \emph {et~al.}(2016)\citenamefont
  {Mesaros}, \citenamefont {Fujita}, \citenamefont {Edkins}, \citenamefont
  {Hamidian}, \citenamefont {Eisaki}, \citenamefont {Uchida}, \citenamefont
  {Davis}, \citenamefont {Lawler},\ and\ \citenamefont {Kim}}]{davis2016b}%
  \BibitemOpen
  \bibfield  {author} {\bibinfo {author} {\bibfnamefont {A.}~\bibnamefont
  {Mesaros}}, \bibinfo {author} {\bibfnamefont {K.}~\bibnamefont {Fujita}},
  \bibinfo {author} {\bibfnamefont {S.~D.}\ \bibnamefont {Edkins}}, \bibinfo
  {author} {\bibfnamefont {M.~H.}\ \bibnamefont {Hamidian}}, \bibinfo {author}
  {\bibfnamefont {H.}~\bibnamefont {Eisaki}}, \bibinfo {author} {\bibfnamefont
  {S.}~\bibnamefont {Uchida}}, \bibinfo {author} {\bibfnamefont {J.~C.~S.}\
  \bibnamefont {Davis}}, \bibinfo {author} {\bibfnamefont {M.~J.}\ \bibnamefont
  {Lawler}}, \ and\ \bibinfo {author} {\bibfnamefont {E.-A.}\ \bibnamefont
  {Kim}},\ }\href@noop {} {\bibfield  {journal} {\bibinfo  {journal} {Proc.
  Nat'l. Acad. Sci.}\ }\textbf {\bibinfo {volume} {113}},\ \bibinfo {pages}
  {12661} (\bibinfo {year} {2016})}\BibitemShut {NoStop}%
\bibitem [{\citenamefont {Xia}\ \emph {et~al.}(2008)\citenamefont {Xia},
  \citenamefont {Schemm}, \citenamefont {Deutscher}, \citenamefont {Kivelson},
  \citenamefont {Bonn}, \citenamefont {Hardy}, \citenamefont {Liang},
  \citenamefont {Siemons}, \citenamefont {Koster}, \citenamefont {Fejer}, ,\
  and\ \citenamefont {Kapitulnik}}]{kapitulnik2008}%
  \BibitemOpen
  \bibfield  {author} {\bibinfo {author} {\bibfnamefont {J.}~\bibnamefont
  {Xia}}, \bibinfo {author} {\bibfnamefont {E.}~\bibnamefont {Schemm}},
  \bibinfo {author} {\bibfnamefont {G.}~\bibnamefont {Deutscher}}, \bibinfo
  {author} {\bibfnamefont {S.~A.}\ \bibnamefont {Kivelson}}, \bibinfo {author}
  {\bibfnamefont {D.~A.}\ \bibnamefont {Bonn}}, \bibinfo {author}
  {\bibfnamefont {W.~N.}\ \bibnamefont {Hardy}}, \bibinfo {author}
  {\bibfnamefont {R.}~\bibnamefont {Liang}}, \bibinfo {author} {\bibfnamefont
  {W.}~\bibnamefont {Siemons}}, \bibinfo {author} {\bibfnamefont
  {G.}~\bibnamefont {Koster}}, \bibinfo {author} {\bibfnamefont {M.~M.}\
  \bibnamefont {Fejer}}, , \ and\ \bibinfo {author} {\bibfnamefont
  {A.}~\bibnamefont {Kapitulnik}},\ }\href@noop {} {\bibfield  {journal}
  {\bibinfo  {journal} {Phys. Rev. Lett.}\ }\textbf {\bibinfo {volume} {100}},\
  \bibinfo {pages} {127002} (\bibinfo {year} {2008})}\BibitemShut {NoStop}%
\bibitem [{\citenamefont {Badoux}\ \emph {et~al.}(2016)\citenamefont {Badoux},
  \citenamefont {Tabis}, \citenamefont {Lalibert\'e}, \citenamefont
  {Grissonnanche}, \citenamefont {Vignolle}, \citenamefont {Vignolles},
  \citenamefont {Béard}, \citenamefont {Bonn}, \citenamefont {Hardy},
  \citenamefont {Liang}, \citenamefont {Doiron-Leyraud}, \citenamefont
  {Taillefer},\ and\ \citenamefont {Proust}}]{taillefer2016}%
  \BibitemOpen
  \bibfield  {author} {\bibinfo {author} {\bibfnamefont {S.}~\bibnamefont
  {Badoux}}, \bibinfo {author} {\bibfnamefont {W.}~\bibnamefont {Tabis}},
  \bibinfo {author} {\bibfnamefont {F.}~\bibnamefont {Lalibert\'e}}, \bibinfo
  {author} {\bibfnamefont {G.}~\bibnamefont {Grissonnanche}}, \bibinfo {author}
  {\bibfnamefont {B.}~\bibnamefont {Vignolle}}, \bibinfo {author}
  {\bibfnamefont {D.}~\bibnamefont {Vignolles}}, \bibinfo {author}
  {\bibfnamefont {J.}~\bibnamefont {Béard}}, \bibinfo {author} {\bibfnamefont
  {D.~A.}\ \bibnamefont {Bonn}}, \bibinfo {author} {\bibfnamefont {W.~N.}\
  \bibnamefont {Hardy}}, \bibinfo {author} {\bibfnamefont {R.}~\bibnamefont
  {Liang}}, \bibinfo {author} {\bibfnamefont {N.}~\bibnamefont
  {Doiron-Leyraud}}, \bibinfo {author} {\bibfnamefont {L.}~\bibnamefont
  {Taillefer}}, \ and\ \bibinfo {author} {\bibfnamefont {C.}~\bibnamefont
  {Proust}},\ }\href@noop {} {\bibfield  {journal} {\bibinfo  {journal}
  {Nature}\ }\textbf {\bibinfo {volume} {531}},\ \bibinfo {pages} {210}
  (\bibinfo {year} {2016})}\BibitemShut {NoStop}%
\bibitem [{\citenamefont {Chern}(2014)}]{chern2014}%
  \BibitemOpen
  \bibfield  {author} {\bibinfo {author} {\bibfnamefont {C.-H.}\ \bibnamefont
  {Chern}},\ }\href@noop {} {\bibfield  {journal} {\bibinfo  {journal} {Annals
  of Physics}\ }\textbf {\bibinfo {volume} {350}},\ \bibinfo {pages} {159}
  (\bibinfo {year} {2014})}\BibitemShut {NoStop}%
\bibitem [{nag()}]{nagaosa2014}%
  \BibitemOpen
  \href@noop {} {}\bibinfo {note} {Private conversation with Naoto Nagaosa in
  2014}\BibitemShut {NoStop}%
\bibitem [{\citenamefont {Lee}\ and\ \citenamefont {Chern}()}]{Lee2018}%
  \BibitemOpen
  \bibfield  {author} {\bibinfo {author} {\bibfnamefont {M.-K.}\ \bibnamefont
  {Lee}}\ and\ \bibinfo {author} {\bibfnamefont {C.-H.}\ \bibnamefont
  {Chern}},\ }\href@noop {} {}\bibinfo {note} {"Theory of Fermi arc in
  cuprates", in preparation.}\BibitemShut {Stop}%
\bibitem [{\citenamefont {Hashimoto}\ \emph {et~al.}(2014)\citenamefont
  {Hashimoto}, \citenamefont {Vishik}, \citenamefont {He}, \citenamefont
  {Devereaux},\ and\ \citenamefont {Shen}}]{shen2014}%
  \BibitemOpen
  \bibfield  {author} {\bibinfo {author} {\bibfnamefont {M.}~\bibnamefont
  {Hashimoto}}, \bibinfo {author} {\bibfnamefont {I.~M.}\ \bibnamefont
  {Vishik}}, \bibinfo {author} {\bibfnamefont {R.-H.}\ \bibnamefont {He}},
  \bibinfo {author} {\bibfnamefont {T.~P.}\ \bibnamefont {Devereaux}}, \ and\
  \bibinfo {author} {\bibfnamefont {Z.-X.}\ \bibnamefont {Shen}},\ }\href@noop
  {} {\bibfield  {journal} {\bibinfo  {journal} {Nature Physics}\ }\textbf
  {\bibinfo {volume} {10}},\ \bibinfo {pages} {483} (\bibinfo {year}
  {2014})}\BibitemShut {NoStop}%
\bibitem [{\citenamefont {Ino}\ \emph {et~al.}(1997)\citenamefont {Ino},
  \citenamefont {Mizokawa}, \citenamefont {Fujimori}, \citenamefont {Tamasaku},
  \citenamefont {Eisaki}, \citenamefont {Uchida}, \citenamefont {Kimura},
  \citenamefont {Sasagawa},\ and\ \citenamefont {Kishio}}]{fujimori1997}%
  \BibitemOpen
  \bibfield  {author} {\bibinfo {author} {\bibfnamefont {A.}~\bibnamefont
  {Ino}}, \bibinfo {author} {\bibfnamefont {T.}~\bibnamefont {Mizokawa}},
  \bibinfo {author} {\bibfnamefont {A.}~\bibnamefont {Fujimori}}, \bibinfo
  {author} {\bibfnamefont {K.}~\bibnamefont {Tamasaku}}, \bibinfo {author}
  {\bibfnamefont {H.}~\bibnamefont {Eisaki}}, \bibinfo {author} {\bibfnamefont
  {S.}~\bibnamefont {Uchida}}, \bibinfo {author} {\bibfnamefont
  {T.}~\bibnamefont {Kimura}}, \bibinfo {author} {\bibfnamefont
  {T.}~\bibnamefont {Sasagawa}}, \ and\ \bibinfo {author} {\bibfnamefont
  {K.}~\bibnamefont {Kishio}},\ }\href@noop {} {\bibfield  {journal} {\bibinfo
  {journal} {Phys. Rev. Lett.}\ }\textbf {\bibinfo {volume} {79}},\ \bibinfo
  {pages} {2101} (\bibinfo {year} {1997})}\BibitemShut {NoStop}%
\bibitem [{\citenamefont {Yagi}\ \emph {et~al.}(2006)\citenamefont {Yagi},
  \citenamefont {Yoshida}, \citenamefont {Fujimori}, \citenamefont {Kohsaka},
  \citenamefont {Misawa}, \citenamefont {Sasagawa}, \citenamefont {Takagi},
  \citenamefont {Azuma},\ and\ \citenamefont {Takano}}]{fujimori2006}%
  \BibitemOpen
  \bibfield  {author} {\bibinfo {author} {\bibfnamefont {H.}~\bibnamefont
  {Yagi}}, \bibinfo {author} {\bibfnamefont {T.}~\bibnamefont {Yoshida}},
  \bibinfo {author} {\bibfnamefont {A.}~\bibnamefont {Fujimori}}, \bibinfo
  {author} {\bibfnamefont {Y.}~\bibnamefont {Kohsaka}}, \bibinfo {author}
  {\bibfnamefont {M.}~\bibnamefont {Misawa}}, \bibinfo {author} {\bibfnamefont
  {T.}~\bibnamefont {Sasagawa}}, \bibinfo {author} {\bibfnamefont
  {H.}~\bibnamefont {Takagi}}, \bibinfo {author} {\bibfnamefont
  {M.}~\bibnamefont {Azuma}}, \ and\ \bibinfo {author} {\bibfnamefont
  {M.}~\bibnamefont {Takano}},\ }\href@noop {} {\bibfield  {journal} {\bibinfo
  {journal} {Phys. Rev. B}\ }\textbf {\bibinfo {volume} {73}},\ \bibinfo
  {pages} {172503} (\bibinfo {year} {2006})}\BibitemShut {NoStop}%
\bibitem [{\citenamefont {Gorny}\ \emph
  {et~al.}(1999{\natexlab{a}})\citenamefont {Gorny}, \citenamefont {Vyaselev},
  \citenamefont {Martindale}, \citenamefont {Nandor}, \citenamefont
  {Pennington}, \citenamefont {Hammel}, \citenamefont {Hults}, \citenamefont
  {Smith}, \citenamefont {Kuhns}, \citenamefont {Reyes},\ and\ \citenamefont
  {Moulton}}]{gorny1999}%
  \BibitemOpen
  \bibfield  {author} {\bibinfo {author} {\bibfnamefont {K.}~\bibnamefont
  {Gorny}}, \bibinfo {author} {\bibfnamefont {O.~M.}\ \bibnamefont {Vyaselev}},
  \bibinfo {author} {\bibfnamefont {J.~A.}\ \bibnamefont {Martindale}},
  \bibinfo {author} {\bibfnamefont {V.~A.}\ \bibnamefont {Nandor}}, \bibinfo
  {author} {\bibfnamefont {C.~H.}\ \bibnamefont {Pennington}}, \bibinfo
  {author} {\bibfnamefont {P.~C.}\ \bibnamefont {Hammel}}, \bibinfo {author}
  {\bibfnamefont {W.~L.}\ \bibnamefont {Hults}}, \bibinfo {author}
  {\bibfnamefont {J.~L.}\ \bibnamefont {Smith}}, \bibinfo {author}
  {\bibfnamefont {P.~L.}\ \bibnamefont {Kuhns}}, \bibinfo {author}
  {\bibfnamefont {A.~P.}\ \bibnamefont {Reyes}}, \ and\ \bibinfo {author}
  {\bibfnamefont {W.~G.}\ \bibnamefont {Moulton}},\ }\href@noop {} {\bibfield
  {journal} {\bibinfo  {journal} {Phys. Rev. Lett.}\ }\textbf {\bibinfo
  {volume} {82}},\ \bibinfo {pages} {177} (\bibinfo {year}
  {1999}{\natexlab{a}})}\BibitemShut {NoStop}%
\bibitem [{\citenamefont {Zheng}\ \emph {et~al.}(1999)\citenamefont {Zheng},
  \citenamefont {Clark}, \citenamefont {Kitaoka}, \citenamefont {Asayama},
  \citenamefont {Kodama}, \citenamefont {Kuhns},\ and\ \citenamefont
  {Moulton}}]{zheng1999}%
  \BibitemOpen
  \bibfield  {author} {\bibinfo {author} {\bibfnamefont {G.~Q.}\ \bibnamefont
  {Zheng}}, \bibinfo {author} {\bibfnamefont {W.~G.}\ \bibnamefont {Clark}},
  \bibinfo {author} {\bibfnamefont {Y.}~\bibnamefont {Kitaoka}}, \bibinfo
  {author} {\bibfnamefont {K.}~\bibnamefont {Asayama}}, \bibinfo {author}
  {\bibfnamefont {Y.}~\bibnamefont {Kodama}}, \bibinfo {author} {\bibfnamefont
  {P.}~\bibnamefont {Kuhns}}, \ and\ \bibinfo {author} {\bibfnamefont {W.~G.}\
  \bibnamefont {Moulton}},\ }\href@noop {} {\bibfield  {journal} {\bibinfo
  {journal} {Phys. Rev. B}\ }\textbf {\bibinfo {volume} {60}},\ \bibinfo
  {pages} {R9947} (\bibinfo {year} {1999})}\BibitemShut {NoStop}%
\bibitem [{\citenamefont {Gorny}\ \emph
  {et~al.}(1999{\natexlab{b}})\citenamefont {Gorny}, \citenamefont {Vyaselev},
  \citenamefont {Pennington}, \citenamefont {Hammel}, \citenamefont {Hults},
  \citenamefont {Smith}, \citenamefont {Baumgartner}, \citenamefont
  {Lemberger}, \citenamefont {Klamut},\ and\ \citenamefont
  {Dabrowski}}]{gorny2001}%
  \BibitemOpen
  \bibfield  {author} {\bibinfo {author} {\bibfnamefont {K.~R.}\ \bibnamefont
  {Gorny}}, \bibinfo {author} {\bibfnamefont {O.~M.}\ \bibnamefont {Vyaselev}},
  \bibinfo {author} {\bibfnamefont {C.~H.}\ \bibnamefont {Pennington}},
  \bibinfo {author} {\bibfnamefont {P.~C.}\ \bibnamefont {Hammel}}, \bibinfo
  {author} {\bibfnamefont {W.~L.}\ \bibnamefont {Hults}}, \bibinfo {author}
  {\bibfnamefont {J.~L.}\ \bibnamefont {Smith}}, \bibinfo {author}
  {\bibfnamefont {J.}~\bibnamefont {Baumgartner}}, \bibinfo {author}
  {\bibfnamefont {T.~R.}\ \bibnamefont {Lemberger}}, \bibinfo {author}
  {\bibfnamefont {P.}~\bibnamefont {Klamut}}, \ and\ \bibinfo {author}
  {\bibfnamefont {B.}~\bibnamefont {Dabrowski}},\ }\href@noop {} {\bibfield
  {journal} {\bibinfo  {journal} {Phys. Rev. B}\ }\textbf {\bibinfo {volume}
  {60}},\ \bibinfo {pages} {R9947} (\bibinfo {year}
  {1999}{\natexlab{b}})}\BibitemShut {NoStop}%
\bibitem [{\citenamefont {Mitrovic}\ \emph {et~al.}(2002)\citenamefont
  {Mitrovic}, \citenamefont {Bachman}, \citenamefont {Halperin}, \citenamefont
  {Reyes}, \citenamefont {Kuhns},\ and\ \citenamefont
  {Moulton}}]{mitrivic2002}%
  \BibitemOpen
  \bibfield  {author} {\bibinfo {author} {\bibfnamefont {V.~F.}\ \bibnamefont
  {Mitrovic}}, \bibinfo {author} {\bibfnamefont {H.~N.}\ \bibnamefont
  {Bachman}}, \bibinfo {author} {\bibfnamefont {W.~P.}\ \bibnamefont
  {Halperin}}, \bibinfo {author} {\bibfnamefont {A.~P.}\ \bibnamefont {Reyes}},
  \bibinfo {author} {\bibfnamefont {P.}~\bibnamefont {Kuhns}}, \ and\ \bibinfo
  {author} {\bibfnamefont {W.~G.}\ \bibnamefont {Moulton}},\ }\href@noop {}
  {\bibfield  {journal} {\bibinfo  {journal} {Phys. Rev. B}\ }\textbf {\bibinfo
  {volume} {66}},\ \bibinfo {pages} {014511} (\bibinfo {year}
  {2002})}\BibitemShut {NoStop}%
\bibitem [{\citenamefont {Zheng}\ \emph {et~al.}(2012)\citenamefont {Zheng},
  \citenamefont {Kawasaki}, \citenamefont {Lin}, \citenamefont {Kuhns},\ and\
  \citenamefont {Reyes}}]{zheng2012}%
  \BibitemOpen
  \bibfield  {author} {\bibinfo {author} {\bibfnamefont {G.~Q.}\ \bibnamefont
  {Zheng}}, \bibinfo {author} {\bibfnamefont {S.}~\bibnamefont {Kawasaki}},
  \bibinfo {author} {\bibfnamefont {C.~T.}\ \bibnamefont {Lin}}, \bibinfo
  {author} {\bibfnamefont {P.~L.}\ \bibnamefont {Kuhns}}, \ and\ \bibinfo
  {author} {\bibfnamefont {A.~P.}\ \bibnamefont {Reyes}},\ }\href@noop {}
  {\bibfield  {journal} {\bibinfo  {journal} {J. Supercond. Nov. Magn.}\
  }\textbf {\bibinfo {volume} {25}},\ \bibinfo {pages} {1249} (\bibinfo {year}
  {2012})}\BibitemShut {NoStop}%
\end{thebibliography}

%

\end{document}